\documentclass[fleqn,usenatbib]{mnras}

\usepackage{newtxtext,newtxmath}

\usepackage[T1]{fontenc}
\usepackage{ae,aecompl}

\usepackage{graphicx}	
\usepackage{amsmath}	
\usepackage{amssymb}	

\newcommand{\sub}[1]{_{\mathrm{#1}}}
\newcommand{\msun}{M$_{\sun}$}
\newcommand{\thegal}{NGC1052-DF2}
\defcitealias{vanDokkum2018_Nature}{vD18}
\defcitealias{Penarrubia2010}{P10}

\def\equationautorefname~#1\null{Eq.~(#1)\null}
\def\figureautorefname~#1\null{Fig.~#1\null}
\title[Tidal stripping and the UDG lacking DM]{Tidal stripping as a possible origin of the ultra diffuse galaxy lacking dark matter}

\author[G. Ogiya]{
Go Ogiya$^{1}$\thanks{E-mail: go.ogiya@oca.eu (GO)}
\\
$^{1}$Laboratoire Lagrange, Universit\'e C\^ote d'Azur, Observatoire de la C\^ote d'Azur, CNRS,\\ \quad Blvd de l'Observatoire, CS 34229, 06304 Nice, France
}

\date{Accepted 2018 July 22. Received 2018 July 2; in original form 2018 April 16.}

\pubyear{2018}

\begin{document}
\label{firstpage}
\pagerange{\pageref{firstpage}--\pageref{lastpage}}
\maketitle

\begin{abstract}
Recent observations revealed a mysterious ultra diffuse galaxy, \thegal{}, in the group of a large elliptical galaxy, NGC1052. Compared to expectations from abundance matching models, the dark matter mass contained in \thegal{} is smaller by a factor of $\sim 400$. We utilize controlled $N$-body simulations of the tidal interaction between NGC1052 and a smaller satellite galaxy, that we suppose as the progenitor of \thegal{}, to test if tidal stripping can explain dark-matter deficiency at such levels. We find that when assuming a tightly bound and quite radial orbit and cored density structure for the dark halo of the satellite, our simulations reproduce well both the mass profile and the effective radius inferred from the observations of \thegal{}. Orbital parameters are in the tail, but still consistent with measurements of their distributions from cosmological simulations. Such strongly dark-matter deficient galaxies, in our scenario, are thus expected to be relatively rare in groups and clusters, and not present in the field.
\end{abstract}



\section{Introduction}
\label{sec:introduction}

A traditional galaxy population, low surface brightness galaxies (LSBs), have been commonly observed as both early (elliptical) and late (disc or irregular) types \citep{Impey1997}. While typical LSBs have sizes of $\sim$\,kpc \citep[e.g.][]{Turner1993,McGaugh1994,Dalcanton1997}, some of them may be larger than 10\,kpc \citep{Sandage1984,Bothun1987}. Recently, \cite{vanDokkum2015} defined a sub-class of LSBs, ultra diffuse galaxies (UDGs). At the distance of the Coma cluster, the UDGs have the luminosity of a dwarf galaxy at the size of a $L\sub{*}$ galaxy. Subsequent studies have followed up their finding in the same galaxy cluster \citep[e.g.][]{Koda2015,Yagi2016}, in other clusters and groups of galaxies \citep[e.g.][]{Mihos2015,vanderBurg2017,Trujillo2017} and even in the field \citep{Papastergis2017}.

While dark matter (DM) is thought to play a key role in galaxy formation \citep{White1978} and typical UDGs are dynamically dominated by DM (e.g. \citealt{vanDokkum2016,Amorisco2018,Sifon2018}), \citet[hereafter \citetalias{vanDokkum2018_Nature}]{vanDokkum2018_Nature} inferred that a unique UDG in the NGC1052 group, \thegal{}, previously identified  by \cite{Fosbury1978} and \cite{Karachentsev2000}, is significantly lacking DM, using the kinetic data of globular clusters in the galaxy.\footnote{Similar approach was adopted by \cite{Toloba2018} to measure the masses of UDGs in the Virgo cluster.} The stellar mass of the galaxy is $2 \times 10^{8}$\,\msun{} and theoretical models based on the abundance matching technique and the standard theoretical framework of galaxy formation \citep[e.g.][]{Behroozi2013, Moster2013} expect that it would be surrounded by a DM halo with a mass of $\sim 5 \times 10^{10}$ \,\msun{}. However, the DM mass of \thegal{} is inferred as only $\sim 10^8$\,\msun{} within $7.6$\,kpc from its centre. Note that the mass distribution of \thegal{} is under debate. \cite{Martin2018} and \cite{Laporte2018} showed that the current observation data only allow to more weakly constrain the mass distribution than what \citetalias{vanDokkum2018_Nature} obtained. \cite{vanDokkum2018_note} updated the observation data and confirmed the results in \citetalias{vanDokkum2018_Nature}. Also note that the distance to \thegal{} may be actually shorter than what \citetalias{vanDokkum2018_Nature} measured and then \thegal{} might be just a typical dwarf galaxy \citep{Trujillo2018}.

Tidal stripping may be a possible mechanism to form such unique DM-deficient galaxies. Stars in a smaller satellite system are tightly bound in the centre of its DM haloes and thus would be more resilient to the tidal forces of larger host systems, compared with the outskirt of the DM halo. For example, \citet[][hereafter \citetalias{Penarrubia2010}]{Penarrubia2010} showed that this mechanism can reduce the dynamical mass-to-light ratio of the satellite systems by a factor of $>100$. 

Motivated by \citetalias{Penarrubia2010}, we investigate using high-resolution $N$-body simulations if tidal stripping can indeed explain the deficiency of DM mass even at the extreme level \citetalias{vanDokkum2018_Nature} obtained. Specifically, we consider the interaction between NGC1052 and a possible progenitor of \thegal{}. In fact, some studies found the presence of HI tidal tails in NGC1052 and suggested that the galaxy may have experienced a merger event with a satellite $\sim$ 1\,Gyr ago \citep{vanGorkom1986, Pierce2005,Fernandez-Ontiveros2011}. \autoref{sec:model} describes the model calibrated for the target galaxy pair. We show the results of our simulations in \autoref{sec:results} and summarize and discuss the results in \autoref{sec:summary}. Throughout this Letter, we use the cosmological parameter set based on \cite{Planck2016} and define the virial radius of DM haloes as enclosing an average density of 200 times the critical density. The virial mass and radius of DM haloes are denoted as $M\sub{200}$ and $r\sub{200}$.

\vspace{-5mm}
\section{Model description} 
\label{sec:model}

\subsection{Structural parameters}
\label{ssec:structure}
In this subsection, we calibrate the structural parameters of the galaxy pair, NGC1052 (host) and a possible progenitor of \thegal{} (satellite), using empirical relations. When the relations require redshift, $z$, or we compute the virial radius of the systems, we set $z=0$. We characterize the initial configuration of the host and satellite with a density profile for spherical systems, 
\begin{eqnarray}
\rho(r) = \frac{\rho\sub{s}}{(r/r\sub{s})^{\alpha}(1+r/r\sub{s})^{\beta-\alpha}}, \label{eq:rho}
\end{eqnarray}
where $\rho(r)$ is the density profile as a function of the distance from the center of the system (or more simply radius). The model has four parameters in total: $\rho\sub{s}$ and $r\sub{s}$ are the scale density and length. The parameters, $\alpha$ and $\beta$, control the inner and outer density slopes, respectively. The concentration of the system is defined as $c \equiv r\sub{200}/r\sub{s}$. For example, the Navarro-Frenk-White (NFW) profile for the density structure of DM haloes \citep{Navarro1997} corresponds to the model of $\alpha=1$ and $\beta=3$. 

First, we use the stellar-to-halo mass relation by \cite{Moster2013}. The stellar mass of NGC1052 is $M\sub{*,h} \approx 10^{11}$\,\msun{} \citep[e.g.][]{Forbes2017} and the relation expects a halo mass of $M\sub{200,h} = 1.1 \times 10^{13}$\,\msun{}. The virial radius of the host halo is $r\sub{200,h}=464$\,kpc. Some observations have measured the dynamical mass enclosed within $\sim 20$\,kpc, corresponding to $\sim 7$ effective radius of the galaxy. \cite{vanGorkom1986} obtained $M(<{\rm 19\,kpc}) = 3.1 \times 10^{11}$\,\msun{} using HI kinematics and \cite{Pierce2005} derived $M(<{\rm 23\,kpc}) = 1.7 \times 10^{12}$\,\msun{} with the kinematic information of globular clusters orbiting around NGC1052. The mass of HI gas is $\sim 5 \times 10^8$\,\msun{} \citep{vanGorkom1986} and that of the central super massive black hole is $\sim 1.5 \times 10^8$\,\msun{} \citep{Woo2002} so that DM is the main contributor in the mass even at the central 20\,kpc. We take only the DM halo into account for simplicity. Assuming the NFW density profile and fixing the enclosed mass, we get $c\sub{h}=5.8$, for the constraint of \cite{vanGorkom1986} and $c\sub{h}=51.2$ for that of \cite{Pierce2005}. Giving the halo virial mass (and redshift), one can get the typical concentration of DM haloes. Since the former is more consistent with the $c(M,z)$ relation by \cite{Ludlow2016}, we adopt it in this Letter. 

For the satellite system, \thegal{}, we adopt the stellar mass measured by \citetalias{vanDokkum2018_Nature}, $M\sub{*,s} = 2 \times 10^{8}$\,\msun{}. The relation by \cite{Moster2013} expects a halo virial mass of $M\sub{200,s} = 4.9 \times 10^{10}$\,\msun{} and the corresponding virial radius is $r\sub{200,s}=77.3$\,kpc. We set the concentration parameter, $c\sub{s}=11.2$, as predicted by the relation of \cite{Ludlow2016}.

Another interesting aspect of the halo structure is the inner density slope. One of the long-standing issues on the standard cold dark matter scenario is the so-called core-cusp problem. While cosmological simulations of cold dark matter model predict the power-law density structure, cusp, at the centre of haloes \citep[e.g.][]{Navarro1997}, the kinetic data of low-mass dwarf galaxies prefer the model with a constant density region, core, in the halo centre \citep[e.g.][]{Moore1994,Burkert1995}.\footnote{The situation is becoming more complicated in recent years. \cite{Oman2015} showed that the rotation curves of observed dwarf galaxies are diverse, i.e. dynamics of some dwarfs can be explained with cuspy profiles while others prefer cored profiles. On the other hand, ones in simulations are given as a function of galaxy mass. The diversity may be described with e.g. the choice of density profile models, deviation from the equilibrium state and angle to observe galaxies \citep[e.g.][]{Read2016}.} The interstellar medium in galaxies heated by supernova feedback temporarily expands, and after losing thermal energy by radiative cooling, it falls back towards the centre again to ignite the next star formation. The cycle of gas expansion and contraction also lead the repetitive change in the gravitational potential and dynamically heat up DM to transform the central cusps into cores \citep[e.g.][]{Governato2010, Pontzen2012, Teyssier2013, Ogiya2014, Onorbe2015, Peirani2017}. \cite{Di_Cintio2014} found an empirical relation to connect the central density structure of DM haloes with the stellar-to-halo mass ratio of galaxies. Applying the relation for our satellite model, we get $\alpha=0.1$. We run simulations varying $\alpha$, but fix $\beta=3$.

We model the stellar component of the satellite with the Hernquist profile \citep{Hernquist1990} which corresponds to the model of $\alpha=1$ and $\beta=4$. Stars are distributed to $r=12.3$\,kpc, determined by the same manner with the definition of the virial radius. We set the concentration for the stellar component, $c\sub{*}=23.0$, that realizes the effective radius of 0.93\,kpc, consistent with the size-mass relation of \cite{Lange2015}.

\vspace{-3mm}
\subsection{Orbital parameters}
\label{ssec:orbit}
We characterize the relative orbit between host and satellite systems with the following two parameters. The first one is the radius of the circular orbit corresponding to the orbital energy, $E$, $r\sub{c}(E)$, scaled by the virial radius of the host, i.e. $x\sub{c} \equiv r\sub{c}(E)/r\sub{200,h}$. The second one is the orbital circularity, $\eta$, defined as $\eta \equiv L(E)/L\sub{c}(E)$, where $L(E)$ and $L\sub{c}(E)$ are the angular momentum of the orbit having the orbital energy, $E$, and that of the circular orbit with the same energy. The distribution of the two parameters have been studied using cosmological $N$-body simulations \citep{Khochfar2006,Wetzel2011,Jiang2015}. While typical orbits have $x\sub{c} \sim 1$ and $\eta \sim 0.5$, we set $x\sub{c} = 0.6$ and $\eta=0.1$, the lower limits of the distributions \citep{vandenBosch2018a}. Thus the orbit we study is more tightly bound by the gravitational potential of the host and more radial compared with the typical ones. The satellite is initially located at the apocentre of the orbit, (559, 0, 0)\,kpc and the velocity vector is (0, 20.6, 0)\,km/s in the host frame.\footnote{The host centre is fixed at the origin throughout simulations.} Here we approximate the two systems as point particles to get the vectors. 

According to \cite{Wetzel2011}, the pericentre of the orbit, $0.003 r\sub{200,h} = 1.4$\,kpc, corresponds to the 1.2th percentile of the distribution. In such a central region of the host galaxy stars, gas disc, and central black hole (that we neglect) all play a role. Thus the tidal force the satellite feels in our model would be weaker than the actual. We emphasize that the goal of this study is to investigate whether tidal stripping can form galaxies with little DM as inferred by \citetalias{vanDokkum2018_Nature} and, in fact, the mechanism would work more significantly.

\vspace{-3mm}
\subsection{Simulation setup}
\label{ssec:setup}
In our simulations, the host is treated as a fixed NFW potential of the parameters described above. The timescale of the orbital decay by dynamical friction \citep{Chandrasekhar1943} would be of the order of $(M\sub{h}/M\sub{s}) \tau\sub{d,h}$ where $\tau\sub{d,h}$ is the free-fall time of the host. Measuring it at $r\sub{200}$, we get $\tau\sub{d,h}=2.96$\,Gyr. The mass ratio between the host and satellite is $M\sub{h}/M\sub{s} \approx 224$ and we can safely neglect dynamical friction because the time scale of the orbital decay would be much longer than the age of the Universe.

We follow the dynamical evolution of the satellite modeled as an $N$-body system that consists of the two spherical components (stars and a DM halo) in the initial configuration and is initially in the dynamical equilibrium state. Previous studies assumed that the satellite systems are always dynamically dominated by DM and neglected (\citetalias{Penarrubia2010}) or modelled the stellar component using mass less particles \citep{Errani2015}. However, the goal of this study is to investigate if tidal stripping can make the satellite galaxy dynamically dominated by stars and thus including the stellar component of the satellite system in a self-consistent way is essential. Setting $\beta \leq 3$, the analytical mass profile of the system diverges in the limit of $r \rightarrow \infty$. To avoid this, we employ the sophistication proposed by \cite{Kazantzidis2006}. The density structure of the system is described as \autoref{eq:rho} at $r \leq r\sub{200}$ and is exponentially decayed beyond $r\sub{200}$. The mass of $0.05M\sub{200}$ is initially distributed beyond $r\sub{200}$, keeping $M(<r\sub{200}) = M\sub{200}$. We numerically compute the phase-space distribution function that takes the potential of the two components into account using the Eddington formula \citep{Eddington1916}. Our satellite system of the two components stays at the initial configuration at least for 10\,Gyr in the absence of the host potential. 

We run two simulations, varying the parameter to control the inner density slope, $\alpha$, of the DM halo of the satellite, for 10\,Gyr. In the the first one, we set $\alpha=0.1$ as predicted by the empirical relation by \cite{Di_Cintio2014} and adopt $\alpha=1.0$, corresponding to the NFW profile, in the second run. 100,761,600 particles (100,352,000 and 409,600 particles are for the DM halo and stellar component, respectively) are employed in each simulation and all particle have the same mass, $\approx 510$\,\msun{}. Numerical computation is performed by a treecode \citep{Barnes1986} designed for Graphics Processing Unit clusters \citep{Ogiya2013}. The opening angle of the tree algorithm is $\theta=0.6$ and the Plummer softening parameter is $\epsilon=0.03$\,kpc, determined by the prescription of \cite{Power2003}. The numerical convergence is verified with simulations, reducing the number of particles by a factor of 10, but fixing the other parameters.

\vspace{-5mm}
\section{Results of simulations}
\label{sec:results}

\begin{figure}
\begin{center}
\includegraphics[width=0.7\columnwidth]{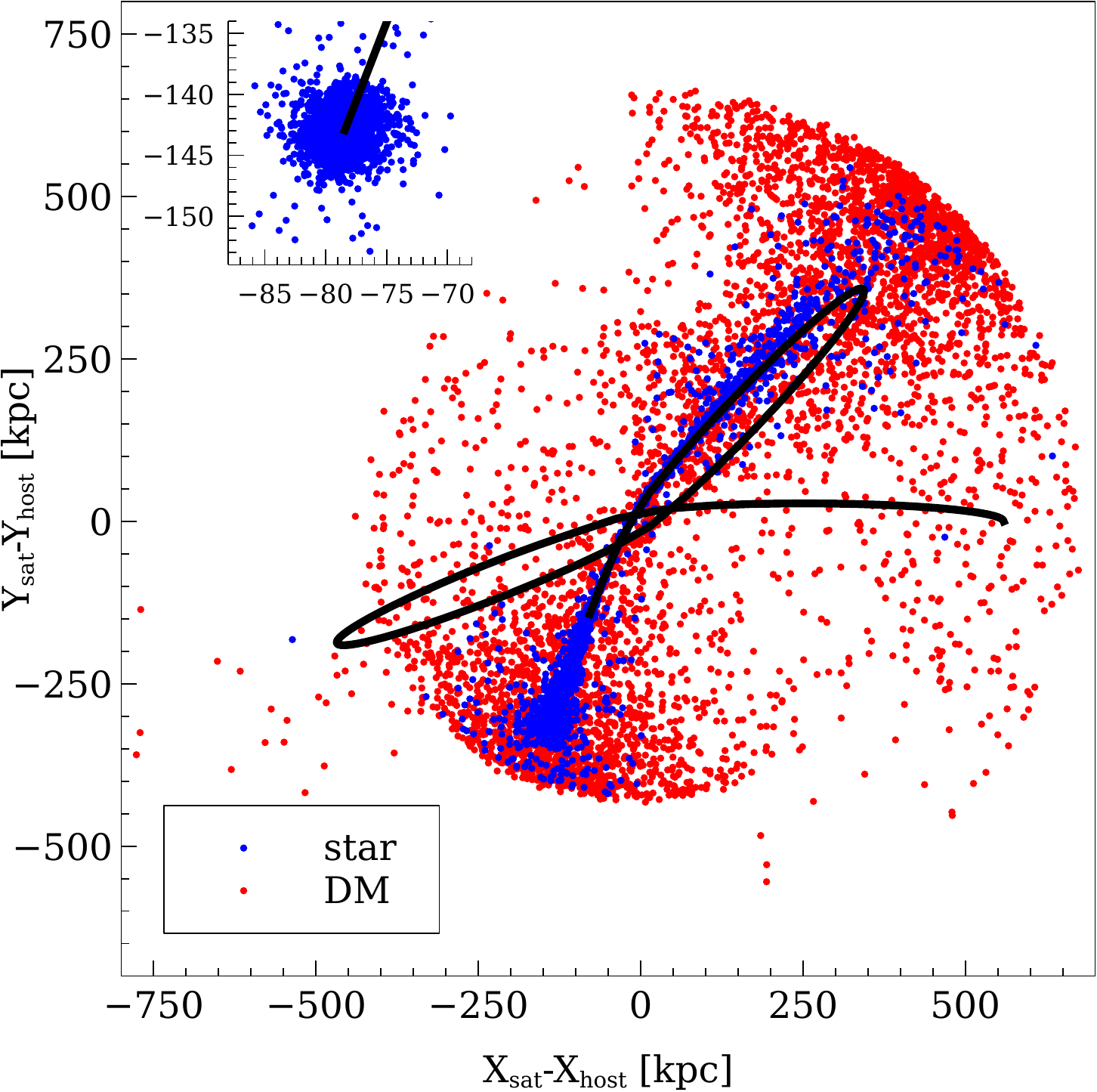}
\end{center}
\vspace{-4mm}
\caption{
Distribution of star (blue) and DM (red) particles in the host frame (the origin corresponds to the centre of the host potential) at $t=10$\,Gyr in the run of $\alpha=0.1$. Black line represents the relative orbit of the satellite system. 5,000 particles are randomly selected from the respective components. The inset panel is a zoomed view of the stellar component that locates at the tip of the orbit, meaning that the stars are settled at the centre of the satellite system. The similar distribution is obtained in the run of $\alpha=1.0$.
\label{fig:dist_map}}
\end{figure}

\autoref{fig:dist_map} illustrates the distribution of stars (blue) and DM (red) at $t=10$\,Gyr in the runs of $\alpha=0.1$. From each component, 5,000 particles are selected and shown. We derive the orbit of the satellite center in the host frame using the scheme described in \cite{vandenBosch2018a}. As demonstrated by many previous studies, the stripped matter remains along the orbit and creates a leading arm and trailing tail, as well as the shell structures at the apocentres. Note that the bulk of the stellar component is still settled at the centre of the satellite (the tip of the black line), as shown in the inset panel.

\begin{figure}
\begin{center}
\includegraphics[width=0.75\columnwidth]{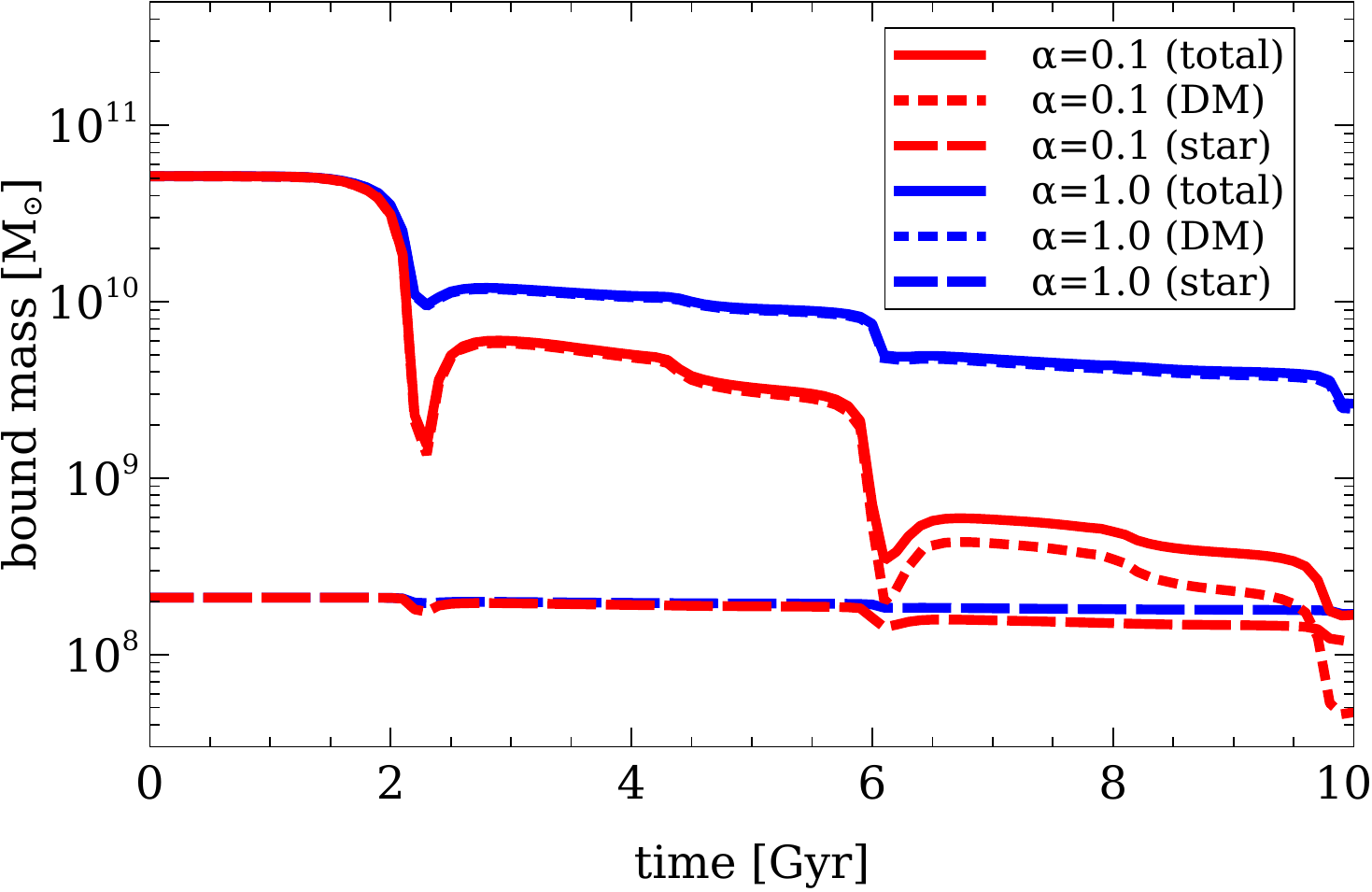}
\end{center}
\vspace{-4mm}
\caption{
Evolution of the bound mass in the satellite systems. Red and blue lines show the results from the runs of $\alpha=0.1$ and 1.0, and solid, dotted and dashed lines represent the total satellite mass, DM and stellar components, respectively. 
\label{fig:evo}}
\end{figure}

In \autoref{fig:evo}, we demonstrate the evolution of the bound mass in the satellite. The stellar mass (dashed) does not significantly change with time in either of the two runs. On the other hand, the bound mass of the DM halo (dotted) decreases with time, especially at the pericentre where the satellite feels the strongest tidal force, and the degree of DM mass loss depends on the inner density slope, $\alpha$ (e.g. \citetalias{Penarrubia2010}; \citealt{Errani2015}). The mass of the cored halo ($\alpha=0.1$; red) decreases by a factor of $\sim 1000$ at $t=10$\,Gyr and, as a result, the stellar component dominates the total mass. In the run of the NFW halo ($\alpha=1.0$; blue), $\sim 5\%$ of the initial DM mass is retained and the system keeps DM dominated as it is in the initial configuration. According to the criteria for numerical convergence in $N$-body simulations of tidal stripping by \cite{vandenBosch2018b}, the derived mass would be reliable in both runs.

\begin{figure}
\begin{center}
\includegraphics[width=0.35\textwidth]{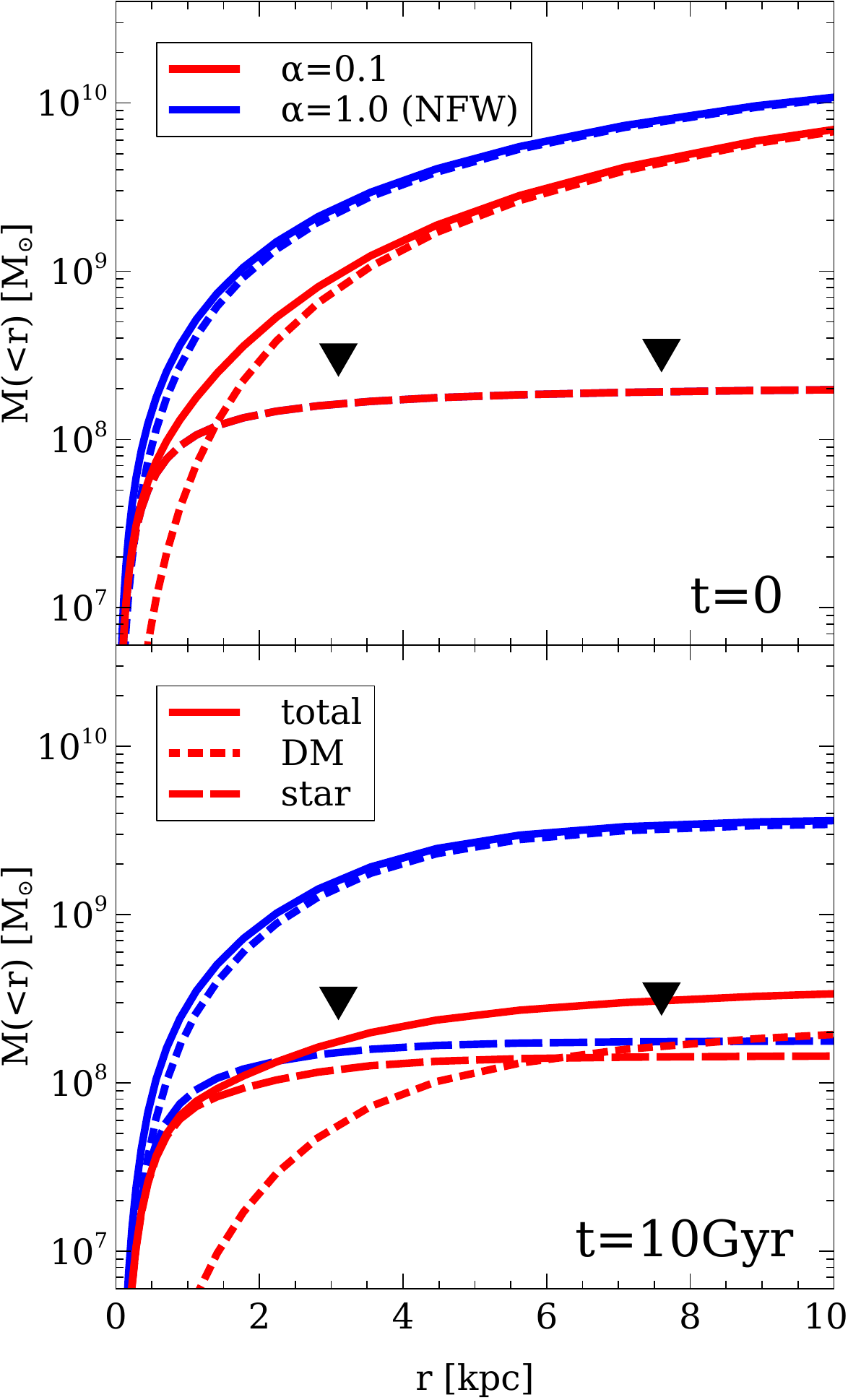}
\end{center}
\vspace{-4mm}
\caption{
Enclosed mass within the radius, $r$. Note that the spherically averaged 3D mass profile is plotted here while the stellar mass profile in \citetalias{vanDokkum2018_Nature} may be obtained from the 2D projected luminosity profile. Upper and lower panels demonstrate the mass profiles at the initial time of simulations and those at $t=10$\,Gyr, respectively. The color and line schemes are the same with those in \autoref{fig:evo}. Black triangles indicate the 90\% confidence upper limits on the enclosed mass of \thegal{} derived by \citetalias{vanDokkum2018_Nature}.
\label{fig:mr}}
\end{figure}

In \autoref{fig:mr}, we compare the mass profiles of the satellite at the initial time of the simulations (upper) and those at $t=10$\,Gyr (lower). The same color and line schemes as in \autoref{fig:evo} are used. Initially, the satellite is a DM dominated system in the both models while the stellar component dominates in mass at $r \la 1.5$\,kpc in the cored model. As implied by \autoref{fig:evo}, the mass profile of the stellar component is hardly altered by the tidal force of the host potential because it is tightly bound. The cored halo model has a smaller enclosed mass than the model of the cuspy NFW halo at the same radius (in the radial range shown in this figure) from the beginning of the simulations. This represents that the cored halo is less tightly bound than the cuspy counterpart. As a result of tidal stripping, the difference in the mass profile is significantly amplified. The cored halo model reproduces the observational result by \citetalias{vanDokkum2018_Nature} well, the stellar and DM mass enclosed within the central 7\,kpc of \thegal{} is $2 \times 10^8$\,\msun{} and $\la 10^8$\,\msun{}.\footnote{Looking at mass contained in each radial bin, DM mass dominates at $r \ga 3$\,kpc in the model of $\alpha=0.1$.} Hence we conclude that tidal stripping by NGC1052 can play an important role in forming a DM-deficient UDG, such as \thegal{}. 

While we also make some tests varying the orbital parameters and the concentration of DM haloes, the parameter range allowed to reproduce the observation is not so large. The distribution of stripped stars would depend on the parameters and may be broader than that \citetalias{vanDokkum2018_Nature} observed. Follow-up deep observations with a wide field of view, e.g. the Hyper Suprime-Cam of Subaru telescope, may help to test the validity of the formation scenario and make better constraints for the parameters. To estimate the surface brightness of the stellar stream in \autoref{fig:dist_map}, we assume the distance to \thegal{} of 20\,Mpc (e.g. \citetalias{vanDokkum2018_Nature}) and stellar mass-to-light ratio of $1 M\sub{\sun}/L\sub{\sun}$. Then we find $>30 \,{\rm mag/arcsec^2}$, below the detection limit of the Dragonfly Telephoto Array. Note that this does not indicate detecting such stellar streams is impossible since the simulation is just an example reproducing the \citetalias{vanDokkum2018_Nature} results and more detailed studies are needed to discuss the distribution of stripped stars and detectability.

\vspace{-5mm}
\section{Summary and discussion}
\label{sec:summary}

We utilize high resolution $N$-body simulations to study the tidal interaction between a large elliptical galaxy, NGC1052, and one of its satellite galaxies and find it can be a possible formation process of the unique DM-deficient UDG, \thegal{} (\citetalias{vanDokkum2018_Nature}). We model an infalling satellite galaxy (the progenitor of \thegal{}) as a two-component (stars and DM) $N$-body system and vary the inner density structure of its DM halo, $\alpha$, motivated by \citetalias{Penarrubia2010}. Other structural parameters and orbital ones are calibrated using the results of the observations, abundance matching techniques as well as numerical simulations. 

Our simulations demonstrate that the observed mass profile of \thegal{} by \citetalias{vanDokkum2018_Nature} is reproduced when i) the DM halo has a central density core; i\hspace{-.1em}i) the satellite is on a tightly bound and quite radial orbit. The orbital parameters are somewhat in the tails of the distributions in cosmological simulations. Because of the rareness of the orbit, our scenario expects that such significantly DM-deficient galaxies would be rare in groups and clusters, and not present in the field.

We also measure the projected effective (half-light) radius, $R\sub{e}$, of the {S{\'e}rsic} profile \citep{Sersic1963}. Stars in the simulations are randomly rotated and the projected radius that contains $50\%$ of the total stellar mass is measured, assuming a two-dimensional axis ratio of $b/a=1$, while \citetalias{vanDokkum2018_Nature} derived $b/a=0.85$, where $a$ and $b$ are the principal semi-axes of the ellipse that satisfy $a \geq b$. We repeat the procedure 100 times and obtain $R\sub{e}= 0.93 \pm 0.00$\,kpc in the initial configuration and $R\sub{e} = 1.97 \pm 0.07$ and $R\sub{e} = 1.17 \pm 0.01$\,kpc at $t=10$\,Gyr in the models of the cored and cuspy profiles. Because the mass in the outskirts of the surrounding DM halo is stripped away, the potential well gets shallower and the stellar component expands as a result. The resultant $R\sub{e}$ in the cored model is approximately consistent with the observed one for \thegal{} (2.2\,kpc). \cite{Carleton2018} used a semi-analytic model based on simulations by \cite{Errani2017} and found similar results. While the mass profile hardly depends on the initial $R\sub{e}$, the resultant $R\sub{e}$ depends on it, according to runs of $N\sim10^7$, varying the initial $R\sub{e}$.

As pointed out by \cite{Martin2018} and \cite{Laporte2018}, large uncertainties remain in estimating the dynamical mass profile of \thegal{} due to the lack of tracers. While stellar streams have not been detected around \thegal{} so far, ones might be found and contribute to provide better mass constraints. More detailed studies would provide predictions for the detectability and distribution of possible stellar streams and test the validity of the formation scenario of \thegal{} considered in this Letter. 

\vspace{-6mm}
\section*{Acknowledgements}
\vspace{-1mm}
GO acknowledges the constructive comments from the referees helped to greatly improve this Letter and thanks Oliver Hahn and Frank van den Bosch for careful reading of the draft. GO was supported by funding from the European Research Council (ERC) under the European Union's Horizon 2020 research and innovation programme (grant agreement No. 679145, project `COSMO-SIMS').



\vspace{-6mm}
\bibliographystyle{mnras}
\bibliography{ti_udg} 






\bsp	
\label{lastpage}
\end{document}